\input epsf
\documentstyle[prl,floats,aps]{revtex}

\begin{document}

\draft

\twocolumn[\hsize\textwidth\columnwidth\hsize\csname @twocolumnfalse\endcsname

\title{Black Hole Spectroscopy:  Determining Waveforms from 3D Excited
Black Holes}

\author{Gabrielle Allen${}^{(1)}$, Karen Camarda${}^{(4)}$, Edward
Seidel${}^{(1,2,3)}$}

\address{
$^{(1)}$Max-Planck-Institut f{\"u}r Gravitationsphysik,
Schlaatzweg 1, 14473 Potsdam, Germany
}
\address{
$^{(2)}$ National Center for Supercomputing Applications,
Beckman Institute, 405 N. Mathews Ave., Urbana, IL 61801
}
\address{
$^{(3)}$ Departments of Physics and Astronomy,
University of Illinois, Urbana, IL 61801
}
\address{
$^{(4)}$ Depts. of Astronomy and Astrophysics, Center for
Gravitational Physics and Geometry,
PSU, University Park, PA 16802
}

\date{\today}
\maketitle

\begin{abstract}
We present the first results for Cauchy nonlinear evolution of 3D,
nonaxisymmetric distorted black holes.  We focus on the extraction and
verification of 3D waveforms determined by numerical relativity.  We
show that the black hole evolution can be accurately followed through
the ringdown period, and comparing with a recently developed
perturbative evolution technique, we show that many waveforms in the
black hole spectrum of modes, such as $\ell=2$ and $\ell=4$, including
weakly excited nonaxisymmetric modes ($m\neq0$), can be accurately
evolved and extracted from the full nonlinear numerical evolution.  We
also identify new physics contained in higher modes, due to nonlinear
effects.  The implications for simulations related to gravitational
wave astronomy are discussed.
\end{abstract}

\vskip2pc]

\narrowtext

Numerical relativity is the main tool for computing waveforms needed
to aid in the detection and interpretation of gravitational wave
signals from merging black holes (BH's).  Recent and very
detailed analysis shows that BH mergers may well be the first events
seen, but that information from full 3D numerical simulations will be
crucial in enhancing the rate of detection and in extracting
astrophysical parameters from the
signals\cite{Flanagan97ab}.

In response to this urgent need for 3D simulations, many groups
worldwide have organized an effort to compute waveforms for binary BH
mergers.  Progress has been slow due to the difficulties of evolving
the full Einstein equations, particularly in the BH case.  For a
review of this difficult problem, see~\cite{Seidel98b}.  Notable
recent progress includes 3D characteristic evolution, success in
apparent horizon boundary conditions (AHBC)\cite{Cook97a,Daues96a},
and extraction of waves from 3D simulations of {\it axisymmetric}
BH's\cite{Camarda97b}.  But, to date there have not been any published
attempts to evolve, and extract meaningful waveforms from true 3D
distorted BH's, as will be formed from the inspiral and merger of two
colliding BH's, using full nonlinear numerical relativity.  This
ability is crucial for gravitational wave astronomy, and in this
Letter we show that it can now be performed with a high degree of
accuracy.

We consider here the first detailed, nonlinear evolutions of
nonaxisymmetric BH's, using the full machinery of standard 3+1
numerical relativity.  We focus on a careful analysis of the 3D
waveforms extracted, with an independent check on their accuracy and
reliability, building particularly on two recent papers.
Ref.~\cite{Camarda97b} showed that one can use a full 3D nonlinear
evolution code to study the excitation and ringdown of axisymmetric
distorted BH's, including the extraction of the waveforms, such as
those created by the head-on collision of two BH's.  It was further
shown, by comparison to 2D codes, that even very low amplitude waves
could be extracted accurately in 3D.  Ref.~\cite{Allen97a} developed a
technique to use perturbation theory to evolve distorted BH data sets
as a testbed for numerical relativity, showing how well this technique
can work on a series of simulations performed with an axisymmetric
distorted BH (2D) code, which allows one to verify the fully nonlinear
code, and to identify nonperturbative effects.

In this paper, for the first time we combine full 3D numerical
relativity with perturbative techniques to show the extent to which
nonaxisymmetric, distorted BH's can be evolved, and waveforms
extracted accurately.  We find that even very low energy ($E <
10^{-6}M$, where $M$ is the BH mass), {\em nonaxisymmetric} wave modes
can be reliably extracted in full 3D simulations.  As emphasized
by\cite{Flanagan97ab}, nonaxisymmetric modes, particularly
the $\ell=m=2$ mode, may be the dominant signal in the final merger
waveform.  We show that not only can this mode be accurately extracted
(albeit for nonrotating spacetimes at present), but even the much
weaker $\ell=4,m$ family can be accurately computed.  For a BH formed
during some violent process (such as collision with another BH), we
expect many modes to be excited.  The main points of this paper are
that (a) entire families of waveforms from the spectrum of an excited
BH can now be reliably determined in full 3D numerical relativity,
verified by perturbation theory, and (b) nonlinear physics can also be
identified.  The combination of 3D nonlinear evolution
and perturbative treatment is very powerful, and provides an important
testing ground for all 3D codes used to study waveforms from BH merger
events, that will be crucial for gravitational wave astronomy.

{\bf 3D Simulations and Comparison with Perturbative Evolution.} We
consider the nonlinear numerical evolution of full 3D distorted BH's
with a code written in 3D Cartesian coordinates, described in
Refs.\cite{Camarda97b,Camarda97a,Camarda97c,Anninos94c,Anninos94d}.
The code is built on earlier 3D work, where
spherical~\cite{Anninos94c} and colliding BH's\cite{Anninos96c}, and
pure waves\cite{Anninos94d} were studied, and was used in
\cite{Camarda97b,Camarda97a,Camarda97c}, where axisymmetric distorted
BH data sets were studied in 3D. Preliminary attempts to extract waveforms from 3D BH's were made 
in~\cite{Camarda97a}. This code solves the full set of
Einstein equations with general gauge and slicing conditions, but here
we use zero shift and the ``1+log'' algebraic lapse
condition\cite{Anninos94c}.

The inner boundary on the BH is provided by an isometry
condition~\cite{Camarda97b,Camarda97a,Camarda97c}.  Ultimately, one
expects to replace this with an AHBC\cite{Seidel98b}, preventing the need
to bend time slices so much that they ultimately destroy the
simulations\cite{Anninos94c}.  The outer boundary conditions keep all
functions fixed in time.  If the boundary is far enough away this
condition is usually adequate, but at present in 3D, without adaptive
meshes (AMR)\cite{Choptuik89,Papadapoulos98a}, it is generally
impractical.  This can create reflections of the waves at late times.
Better boundary conditions, such as the perturbative outer
boundary\cite{Abrahams97a}, Cauchy-Characteristic
matching\cite{Bishop98a}, other treatments provided by hyperbolic
formulations of Einstein's equations, or AMR should improve this
situation.  But these issues do not affect the main point we make
below, which is that {\em high resolution and verifiable studies of
truly 3D BH's, including the ringdown and propagation of very low
amplitude nonaxisymmetric radiation modes, are now possible.} 

The initial data we evolve in this paper belong to the family of
distorted BH's discussed in
Refs.\cite{Camarda97b,Camarda97a,Camarda97c,Abrahams92a,Bernstein94a,Brandt94a},
and correspond to a ``Brill wave''\cite{Brill59} superimposed on a BH.
Such data sets mimic the endstate of two BH's colliding, forming a
useful model for studying the late stages of BH mergers.  To summarize
we simply write the 3--metric here as $d\ell^2 = \tilde{\psi}^4 \left(
e^{2q} \left( d\eta^2 + d\theta^2 \right) + \sin^2\theta d\phi^2
\right), $where $\eta$ is a radial coordinate related to the Cartesian
coordinates by $\sqrt{x^{2}+y^{2}+z^{2}} = e^{\eta}$\cite{Camarda97b}.
Given a choice for the ``Brill wave'' function $q$, the Hamiltonian
constraint leads to an elliptic equation for the conformal factor
$\tilde{\psi}$.  The function $q$ represents the wave surrounding the
BH, and is chosen to be $ q\left(\eta,\theta,\phi\right) = a
\sin^n\theta \left( e^{-\left(\frac{\eta+b}{w}\right)^2} +
e^{-\left(\frac{\eta-b}{w}\right)^2} \right) \left(1+c
\cos^2\phi\right).$ 
If the amplitude $a$ vanishes,
the undistorted Schwarzschild solution results; small values
of $a$ correspond to a perturbed BH. Although these data
sets can include angular momentum \cite{Brandt97a}, we focus here on
the time symmetric case, so only the Hamiltonian constraint need be
solved for the conformal factor $\tilde{\psi}$.  Note that the form of
the 3--metric, although fully nonaxisymmetric, does have a discrete
(quadrant) symmetry about the $z-$axis and the equatorial plane.
Hence we can evolve the BH in an octant and save on the memory and
computation required.

{\em Full 3D Nonlinear Evolution.} The initial data set
$(a=-0.1,b=0,c=0.5,w=1,n=4)$ was evolved with the 3D 
code described above with a grid spacing of
$\Delta x = \Delta y = \Delta z = 0.093M$, with $300$ grid zones
in each coordinate direction.  As in previous spherical and
axisymmetric studies, the evolutions can be carried out through about
$t=40M$, when the large gradients in metric functions, 
created by the use of singularity avoiding time slicings,
cause the code to become rather inaccurate and crash.  The problems
encountered, not relevant for the main results of this work, 
should be
eliminated through the use of AHBC (see, e.g.\cite{Seidel98b} and
references therein).

During the evolution, we extract the waves using a gauge-invariant
technique developed originally by
Abrahams\cite{Abrahams88,Abrahams90}.  The details of the present 3D
application are given in~\cite{Allen97a,Camarda97a,Camarda97c}.
Essentially, the Zerilli function $\psi$ is computed at various radii
away from the distorted BH. In this work we extracted waveforms at
Schwarzschild radii, $r$, of $8.0M$, $10.2M$, $12.6M$ and $14.9M$.  The 3D
code and waveform extraction procedure were developed and verified
through careful comparison to an axisymmetric
code\cite{Abrahams92a,Bernstein94a} and also on full 3D initial
data\cite{Allen97a}.  Here we apply the same techniques to study the
first waveforms emitted from {\it nonaxisymmetric} distorted BH's in
full numerical relativity.

\begin{figure}
\epsfxsize=250pt \epsfbox{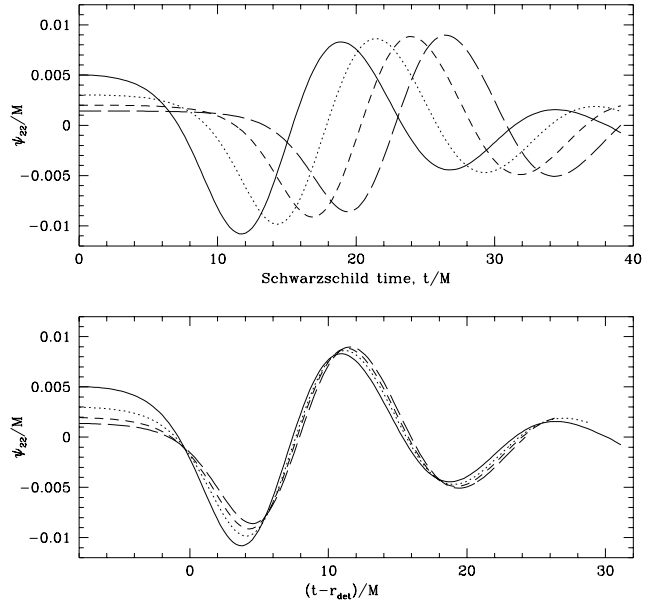}
\caption{\label{wave22}
We show waveforms for the $\ell=m=2$ nonaxisymmetric mode extracted
from the full nonlinear 3D simulation of the distorted BH. The top
graph shows the waveform extracted at different locations, while the
bottom shows the phase-shifted waveforms, allowing comparison between
them.}
\end{figure}

In Fig.~\ref{wave22} we show the nonaxisymmetric $\ell=m=2$ mode
emitted during the evolution of this distorted BH. Waves are
extracted at the different radii, showing the development of this mode
in time.  The quasinormal ringing of the BH is clearly evident and has
the correct complex frequency. 
The modes are normalized so that the energy they carry is given by
$\dot{E} = (1/32\pi) \dot{\psi}^2$.  For this mode, we find that the
energy is about $10^{-6}M$.

As mentioned, this 3D mode is expected to be an important and dominant
mode in the spectrum of BH excitations created during the merger
process.  This is the first example showing that nonaxisymmetric modes
can be computed in a full 3D simulation, in spite of all the
difficulties of BH simulations. The calculation was repeated at
higher and lower grid resolutions, the waveform was found to be converging
to between first and second order in the grid spacing, with the
changes in the waveform at different resolutions being minimal.
Beyond resolution studies, having an independent technique
to determine the expected waveform, to which comparisons can be made,
is also essential in studying their quality.  As emphasized
elsewhere\cite{Pullin98a}, perturbative techniques can also be used as
a testbed to study black hole dynamics, and provide physical insight
into the nature of the solution.  We turn to perturbation theory in
the next section to provide this testbed, and the physical insight it
brings.

{\em Comparison with Perturbative Evolution.} Although we have seen
that the 3D modes have the right quasinormal frequency, without a
careful check of the simulations, we have no guarantee that the
results are actually correct.  However, building on prior work of
Abrahams, Price, Pullin and others (see \cite{Pullin98a} for recent
review), we have recently developed a technique to study distorted
BH's through a perturbative approach to the evolution\cite{Allen97a}.
The technique makes use of the extraction technique used above to
extract initial data for the Zerilli equation, describing linearized
perturbations of spherical BH's.  As long as the initial wave content
on the Schwarzschild background is small, one should be able to
use this simple linear wave equation to evolve the gravitational waves
and predict the waveforms that must be seen by the full 3D, nonlinear
code.  We have taken such initial data, evolved it with the Zerilli
equation, and compared with waveforms extracted from the full 3D
nonlinear evolution in Cartesian coordinates.
\begin{figure}
\epsfxsize=250pt \epsfbox{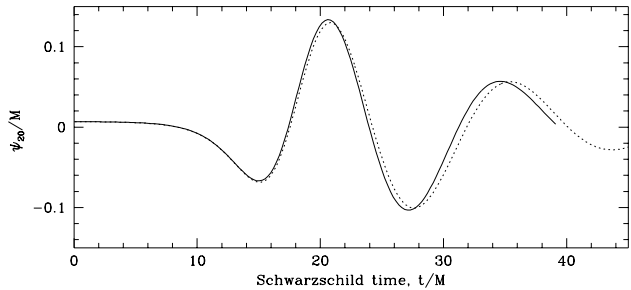}
\caption{\label{wave20}
We show the waveform for the $\ell=2,m=0$ mode,
extracted from the linear and nonlinear evolution codes.  The dotted
(solid) line shows the linear (nonlinear) evolution.}
\end{figure}

\begin{figure}
\epsfxsize=250pt \epsfbox{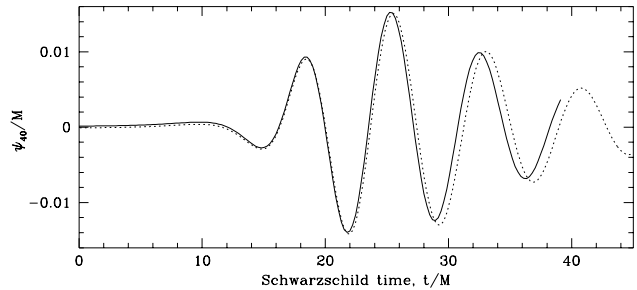}
\caption{\label{wave40}
Waveforms are shown for the $\ell=4$, $m=0$  extracted from the
linear and nonlinear evolution codes. The dotted (solid) line shows
the linear (nonlinear) evolution.
}
\end{figure}

\begin{figure}
\epsfxsize=250pt \epsfbox{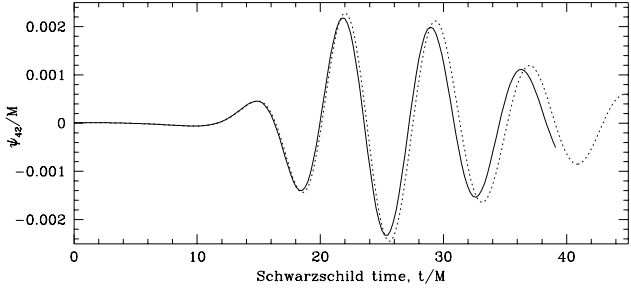}
\caption{\label{wave42}
Waveforms are shown for the $\ell=4,m=2$ mode, extracted from the
linear and nonlinear evolution codes.  The dotted (solid) line shows
the linear (nonlinear) evolution.  }
\end{figure}

In Figs.~\ref{wave20}, \ref{wave40}, and \ref{wave42} we show
results for a large part of the BH spectrum of modes excited.  The
waveforms for the linear and nonlinear evolutions are each plotted on
the same graphs, extracted at $r=12.6M$.  The total energy radiated
for these modes runs from $E\sim 3\times 10^{-4}M$ for the $\ell=2$, $m=0$
mode, to $ E\sim 3\times 10^{-7}M$ for the $\ell=4$, $m=2$ mode.  The
agreement between these two completely independent treatments is
remarkable, giving complete confidence in the reliability of these
waveforms.  Not only do the perturbative results confirm those of full
3D numerical relativity, but the 3D results confirm the perturbative
treatment.  This is an important point to emphasize, as it shows that
these modes are actually in the perturbative regime.  Hence the
gravitational wave physics of these modes is adequately treated by
linearized theory.  This must be captured by the full 3D simulation,
even though the singularity avoiding slicing chosen for the 3D
simulation forces the BH background to evolve in a fully nonlinear
manner (leading to ``grid stretching'', etc.)  The $\ell=m=2$ waveform
already shown above has the same level of agreement with its
perturbative counterpart.  A full analysis of these waveforms and
other 3D BH spacetimes will be published elsewhere\cite{Allen98b}.

We emphasize that although the amplitude of distortion of this BH is
low enough that some modes can be treated with linear perturbation
theory, there are limits to its applicability.  Other modes can be out
of the linear regime, and nonlinear evolution should not agree with
linearized evolution results.  Detailed analysis of the initial data
shows that only the modes $\ell=2$, $m=0,2$ and $\ell=4$, $m=0,2$
should occur at linear order in the wave amplitude $a$, all other
modes occurring at order $a^2$ or above.  These higher order modes
cannot be accurately evolved with the first order Zerilli equation;
Ref.~\cite{Gleiser96b} shows that in such cases there is a source term
arising from {\em nonlinear} mixing of the modes that appear at linear
order.  In some sense, these modes are {\em too small} for linear
treatment to adequately capture the physics, yet as we show now they
can be treated in full 3D numerical relativity.

\begin{figure}
\epsfxsize=250pt \epsfbox{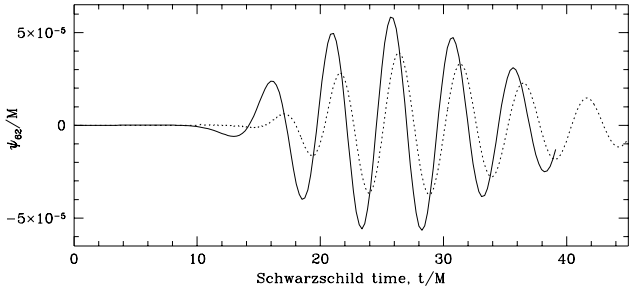}
\caption{\label{wave62}
Waveforms are shown for the $\ell=6,m=2$ mode, extracted from the
linear and nonlinear evolution codes.  The dotted (solid) line shows
the linear (nonlinear) evolution.  The discrepancy is attributed to a
{\em nonlinear} effect.}
\end{figure}

In Fig.~\ref{wave62}, we consider such a low amplitude case, $l=6$,
$m=2$. (Other modes are more difficult to extract because of numerical
resolution, e.g. $\ell=m=4$ has not converged in our
simulations.  Full details will be published elsewhere).  This is a
 difficult mode to extract numerically in 3D~\cite{Allen98b}, but
the waveform is not significantly altered by more resolution,
and show a marked disagreement in contrast to the previous Figs.  
The linear energy calculation indicates that the
energy of this mode is only $E\sim 10^{-10}M$.  
Simulations at different values of $a$ show the amplitude of
this waveform scaling as $a^2$.  Hence, we fully expect the
perturbative approach to fail for this mode, and we must turn to
nonlinear treatments, such as a 3D numerical relativity code, to
compute it.  This same effect is discussed in~\cite{Allen97a} for
highly resolved 2D cases where nonlinear mode mixing is very clearly
seen; details of the more difficult 3D cases are in preparation for
publication elsewhere\cite{Allen98b}.


{\bf Conclusions.} We have shown for the first time that evolutions of
a new class of full 3D distorted black holes can be accurately
performed in 3D Cauchy nonlinear evolution codes in Cartesian
coordinates, and further that a large spectrum of different {\em
  nonaxisymmetric} waveforms can now be determined from the simulations,
even when they carry a tiny fraction of the ADM mass ($E\sim
10^{-6}M$).  Such capability will be crucial in simulating 3D black
hole mergers, needed for effective detection of waves.  We have also
applied newly developed perturbative techniques to verify that
the waveforms are very accurate, and also to show {\em nonlinear}
effects.  This comparison technique should provide a powerful testbed
and analysis tool for all 3D codes under development for use in
gravitational wave astronomy.

{\bf Acknowledgments.} This work was supported by AEI, NCSA, and NSF
PHY/ASC 9318152 (ARPA supplemented).  We thank K.V. Rao and John Shalf
for assistance with the computations, and many colleagues at NCSA,
AEI, and Washington University, especially M. Alcubierre,  B. Br\"{u}gmann, 
C. Gundlach and J. Mass\'{o}, who have influenced this work. 
E.S. would like to thank the University of the Balearic Islands 
for their hospitality.
Calculations were performed at AEI and NCSA on SGI/Cray Origin 2000
supercomputers.


\end{document}